\documentclass[10pt, conference, compsocconf]{IEEEtran}
\pdfoutput=1
\usepackage{cite}
\usepackage[cmex10]{amsmath}
\usepackage{amssymb,amsfonts}
\usepackage{algorithmic}
\usepackage{graphicx}
\usepackage{textcomp}
\usepackage{xcolor}
\usepackage[hyphens]{url}

\def\BibTeX{{\rm B\kern-.05em{\sc i\kern-.025em b}\kern-.08em
    T\kern-.1667em\lower.7ex\hbox{E}\kern-.125emX}}

\usepackage{braket}
\usepackage{bm}
\usepackage{mfirstuc}
\usepackage{physics}
\usepackage{caption}
\usepackage{subcaption}
\usepackage[frozencache,cachedir=cache/]{minted}

\begin{document}
%
\title{SupermarQ: A Scalable Quantum Benchmark Suite}




%
\author{
\IEEEauthorblockN{
Teague Tomesh\IEEEauthorrefmark{1}\IEEEauthorrefmark{2}\IEEEauthorrefmark{5},
Pranav Gokhale\IEEEauthorrefmark{2},
Victory Omole\IEEEauthorrefmark{2},
Gokul Subramanian Ravi\IEEEauthorrefmark{3},
Kaitlin N. Smith\IEEEauthorrefmark{3}, \\
Joshua Viszlai\IEEEauthorrefmark{3},
Xin-Chuan Wu\IEEEauthorrefmark{3}\IEEEauthorrefmark{6},
Nikos Hardavellas\IEEEauthorrefmark{4},
Margaret R. Martonosi\IEEEauthorrefmark{1} and
Frederic T. Chong\IEEEauthorrefmark{2}\IEEEauthorrefmark{3}}
\IEEEauthorblockA{\IEEEauthorrefmark{1}Department of Computer Science, Princeton University; \IEEEauthorrefmark{2}Super.tech; \IEEEauthorrefmark{3}Department of Computer Science,\\ University of Chicago; \IEEEauthorrefmark{4}Department of Computer Science, Northwestern University; \IEEEauthorrefmark{6}Intel Labs, Intel Corporation}
\IEEEauthorblockA{\IEEEauthorrefmark{5}Email correspondence: ttomesh@princeton.edu}}


\maketitle

\begin{abstract}
The emergence of quantum computers as a new computational paradigm has been accompanied by speculation concerning the scope and timeline of their anticipated revolutionary changes. While quantum computing is still in its infancy, the variety of different architectures used to implement quantum computations make it difficult to reliably measure and compare performance. This problem motivates our introduction of SupermarQ, a scalable, hardware-agnostic quantum benchmark suite which uses application-level metrics to measure performance. SupermarQ is the first attempt to systematically apply techniques from classical benchmarking methodology to the quantum domain. We define a set of feature vectors to quantify coverage, select applications from a variety of domains to ensure the suite is representative of real workloads, and collect benchmark results from the IBM, IonQ, and AQT@LBNL platforms. Looking forward, we envision that quantum benchmarking will encompass a large cross-community effort built on open source, constantly evolving benchmark suites. We introduce SupermarQ as an important step in this direction.
\end{abstract}


%

\section{Introduction}
The creation, validation, and implementation of benchmarks is a foundational aspect of computer architecture. The pursuit of increasingly powerful computers has resulted in a zoo of computational architectures which requires the use of application benchmarks to enable sensible, cross-platform performance measurements.

The emergence of new computational paradigms motivates the development and deployment of new benchmark suites to measure and define performance. The upsurge of computing in the 1970s and 80s led to the creation of LINPACK and SPEC for benchmarking supercomputers and workstations~\cite{dongarra2003linpack, henning2006spec}. The PARSEC benchmark suite was introduced in response to the proliferation of chip multi-processors~\cite{bienia2008parsec}, and the explosion of interest in machine learning applications led to the creation of MLPerf to benchmark performance between different models~\cite{mattson2020mlperf}. Similarly, the emergence of new quantum computer architectures must be matched by the development of a new suite of benchmarks tailored to these systems.

Prior attempts to benchmark quantum processors have focused on single-number metrics to quantify performance. For example, the quantum volume~\cite{cross2019validating} and Q-score~\cite{martiel2021benchmarking} metrics target a specific class of circuits or a single application, respectively, to determine the overall performance of a quantum processing unit (QPU). However, capturing the general performance of a computational system within a single number can be very challenging as well as misleading. Throughout the history of classical benchmarking there have been examples of compilers and microarchitectures optimized for specific benchmarks while neglecting the application domains that fall outside the scope of the benchmark suite~\cite{hennessy2011computer}. Therefore, it is advantageous to use an entire suite of benchmarks to obtain a better sense of system performance across a range of potential applications.

Application-level benchmarks provide more accurate measurements of system-level performance than circuit- and gate-level strategies which are better suited to characterizing specific properties of the hardware. Applications also differ in the amount and kind of resources they require. Therefore, a benchmark suite must maintain good coverage of the application space to accurately represent realistic workloads. We introduce a set of feature vectors to describe and measure the coverage of quantum applications. Each benchmark application is described by a single vector, and the individual features that make up this vector are based on hardware-agnostic quantities that are related to the application's resource requirements.

Existing quantum processors are described as Noisy Intermediate-Scale Quantum (NISQ) devices due to their prohibitive gate error rates and limited number of qubits~\cite{preskill2018quantum}. NISQ computers lack the computational resources to run the originally-envisioned quantum applications such as factoring~\cite{shor1999polynomial}, database search~\cite{grover1996fast}, and solving linear systems~\cite{harrow2009quantum}; which require devices that are fault-tolerant (FT). A quantum benchmark suite must take into account the gap between the machines of today and those of tomorrow by incorporating applications that scale \textit{down} to the NISQ and \textit{up} to the FT regime in order to remain relevant.

The state-of-the-art in quantum computing is rapidly progressing. As qubit counts increase and gate errors decrease, new use cases may be discovered. The set of benchmark applications should change to reflect those developments. In addition, quantum software techniques are continuously improving and adapting to changes in hardware. This aspect of quantum computing should be reflected in the benchmark suite by evaluating the performance of the system, composed of the hardware and the software, as a whole. Some compiler optimizations, such as noise-aware qubit placement~\cite{ash2019qure, bhattacharjee2019muqut, murali2019noise, tannu2019not}, have already become standard practice within some quantum compilation toolflows and can make the difference between program success and failure.

Recent works within the quantum computer architecture community have taken the first steps towards quantum benchmarking. The PPL+2020~\cite{patel2020experimental} suite was evaluated on seven superconducting QPUs, focused on characterizing the error rates of different operations, and demonstrated the time dependence of their performance. The TriQ~\cite{murali2019full} suite was used to perform a cross-platform comparison between superconducting and trapped-ion systems and revealed the importance of software visibility into the hardware's native gates. However, the scalability of these suites is limited by their reliance on circuit simulation to estimate how well the QPUs are performing. SupermarQ extends these works by introducing a systematic and principled approach to building a scalable quantum benchmark suite.
We introduce a set of principles: (1) scalability, (2) meaningful and diverse applications, (3) full-system evaluation, and (4) adaptivity, to address the constraints presented above and provide a basis for developing a robust suite of benchmarks.

Resources such as coherence time, the number of qubits, and number of two-qubit gates required by a quantum program significantly impact that program's success rate~\cite{patel2020experimental, murali2019full}.
We introduce multiple features including the connectivity of the logical circuit, the degree of parallelism, and the proportion of two-qubit entangling operations within the circuit to reflect an application's resource requirements. We use these features to examine the coverage of existing quantum benchmark suites, and given a quantum device and benchmark application, we study the correlation between the application's features and the performance of the QPU.

We seek to define the challenges that surround the construction of a scalable quantum benchmark suite and meet these challenges by drawing on techniques from classical benchmarking. To this end, our contributions include:
\begin{itemize}
    \item A set of guiding principles that define the desirable qualities of a scalable quantum benchmark suite.
    \item A set of feature vectors to characterize the applications and coverage of quantum benchmark suites.
    \item The discovery that realistic benchmark suites give better coverage than existing single-application benchmarks and synthetic suites that focus on individual features.
    \item Eight benchmark applications; specified at the level of OpenQASM~\cite{cross2017open} that consist of an open-source circuit generator and performance metric that are both scalable.
    \item Cross-platform evaluation on superconducting and trapped ion architectures.
    \item Correlation of the application features with the observed system performance.
\end{itemize}

The remainder of the paper is organized as follows: we begin with an overview of prior quantum benchmarks in Sec.~\ref{sec:prior work}. In Sec.~\ref{sec:design} we describe the design choices behind the benchmark principles and feature vectors. The benchmark applications and the coverage of different benchmark suites are discussed in Sec.~\ref{sec:benchmarks}. We then step through our methodology in Sec.~\ref{sec:methodology} and evaluate our results in Sec.~\ref{sec:results}. Finally, we provide a discussion of these results in Sec~\ref{sec:discussion} and close with final remarks and future work in Sec.~\ref{sec:conclusion}.

\section{Prior Work}\label{sec:prior work}

\subsection{Classical Benchmarks}
As processing power grew exponentially with Moore's Law it was necessary for the development of classical benchmark suites to keep pace so that the performance of newly emerging architectures could be accurately measured. Advancements in areas such as high-performance computing, workstations, chip multi-processors, and machine learning were accompanied by new suites of benchmarks designed to quantify performance within each respective domain~\cite{dongarra2003linpack, henning2006spec, bienia2008parsec, mattson2020mlperf}. 

In particular, the PARSEC benchmark suite was designed around a set of principles that helped define its scope and purpose. The five requirements that PARSEC aimed to meet were: the inclusion of multithreaded applications, representing emerging workloads, targeting diverse workloads, utilizing state-of-art techniques, and supporting on-going research efforts~\cite{bienia2008parsec}. SupermarQ is inspired by the principled approach taken by PARSEC because of the similarities between the emergence of chip multi-processors and the emergence of quantum computers.

\subsection{Quantum Benchmarks}
The current state of quantum benchmarks consist of (a) low-level approaches to measuring individual gate errors, qubit coherence times, or other hardware-level properties, (b) synthetic benchmarks that utilize random circuits to measure hardware performance, (c) single application benchmarks that focus on a particular use-case, and (d) a few examples of initial quantum benchmark suites. Each of these approaches have advanced the state-of-the-art in quantum benchmarking. In the following sections we discuss the tradeoffs associated with each approach.

\subsubsection{Gate-Level Characterization}
The original motivation behind the development of quantum benchmarks was the desire to understand exactly what process the quantum hardware was implementing in the presence of imperfect controls and noise. Quantum process tomography is a well-known technique which can be used to fully characterize any quantum process~\cite{chuang1997prescription}. Unfortunately, this technique scales exponentially with the number of qubits and is therefore only applicable to systems of only a few qubits. In response to the intractability of quantum process tomography, randomized approaches to quantum benchmarking were introduced~\cite{magesan2011scalable, magesan2012characterizing, proctor2019direct}. These methods scale polynomially with the number of qubits and can be used to characterize the average error rates for the different operations within a QPU's native gate set. While understanding the error rates of individual gate operations is a critical component of designing a QC system, especially for constructing noise models, it does not directly capture how the system will perform on real-world applications.

\subsubsection{Synthetic Benchmarks}
Synthetic benchmarks such as the quantum volume protocol~\cite{cross2019validating} and quantum LINPACK benchmark~\cite{dong2020random} have also been introduced to measure the performance of QC systems. Both benchmarks rely on some aspect of randomness within their protocol. The quantum volume metric is computed by finding the largest random circuit of equal width and depth that a QPU is able to execute while generating the correct outputs with probability greater than $2/3$ (i.e., heavy-output generation)~\cite{cross2019validating}. The quantum LINPACK benchmark is inspired by the classical LINPACK benchmark which measures performance by a computer's ability to solve random systems of linear equations.

The main drawbacks to these synthetic benchmarks is that they are neither meaningful nor scalable. Typical quantum applications do not generally take the form of random quantum circuits and therefore the quantum volume and LINPACK benchmarks are not necessarily representative of useful workloads~\cite{preskill2018quantum}. In addition, the computation required to verify the output of these benchmarks becomes intractable as the number of qubits increases. The quantum volume metric requires that the heavy-outputs of the random circuit be computed beforehand, using a classical technique which scales exponentially with the number of qubits~\cite{cross2019validating}. Verification of the quantum LINPACK benchmark also scales unfavorably. In fact, the hardness of this benchmark is based on the same type of chaotic quantum evolution that underlies prior supremacy experiments~\cite{dong2020random, arute2019quantum}. Although quantum LINPACK may be a suitable candidate for testing quantum supremacy, this characteristic is not desirable as a scalable quantum benchmark.

\subsubsection{Application Benchmarks}
The Variational Quantum Eigensolver (VQE)\cite{peruzzo2014variational} is a hybrid quantum-classical algorithm used to compute molecular ground state energies and has been proposed as a potential quantum benchmark~\cite{mccaskey2019quantum}. The \textit{effective fermionic length} is another benchmark which uses VQE to compute the ground state energies of one-dimensional Fermi Hubbard models of increasing length~\cite{dallaire2020application}. 

The Quantum Approximate Optimization Algorithm (QAOA) \cite{farhi2014quantum} has also been proposed as an effective application benchmark. The performance of QAOA on superconducting QPUs was compared against the D-Wave 2000Q quantum annealer for instances of weighted MaxCut and 2-SAT problems~\cite{willsch2020benchmarking}. Another example, the ``Q-score'' performance metric, is computed by finding the largest MaxCut instance which a QPU can effectively solve~\cite{martiel2021benchmarking}.

All of these application based benchmarks possess a level of scalability that is not present in the low-level and synthetic benchmarks. This is due to their use of application-level metrics, like ground state energy or approximation ratio to measure performance. Simultaneously, reliance on application-level metrics makes cross-platform comparisons between different quantum architectures and classical approaches straightforward. This is important because the crossover point between the best classical and quantum approaches is a constantly moving target that shifts with every advance in algorithms, software, and hardware.

Despite the scalability offered by these application benchmarks, a single application is inadequate for measuring overall system performance. Many different applications are required to reflect the diversity of possible workloads.

\subsubsection{Benchmark Suites}
Some prior works have begun to explore the creation of quantum benchmark suites to enable more accurate characterizations of system performance and cross-platform comparisons. QASMBench~\cite{li2020qasmbench} is a low-level benchmark suite based on the OpenQASM assembly language~\cite{cross2017open}. PPL+2020 evaluated nine benchmarks on seven different IBM superconducting QPUs, characterizing their error rates and performance over time~\cite{patel2020experimental}. While both are examples of early quantum benchmark suites, their performance metrics are based on comparisons between the experimental and ideal circuit outputs. This limits the scalability of these suites due to the exponential scaling of quantum circuit simulation.

The current QC landscape is filled with a variety of architectures such as photonic, trapped ion, and superconducting implementations. Initial architectural comparisons between these implementations have revealed the impact that qubit connectivity, native gate operations, and error rates can have on program execution~\cite{murali2019full, blinov2021comparison}. Thus far, however, these cross-platform comparisons have been limited to a handful of applications that do not always represent the workloads we expect to run on QPUs in the near future.

\section{Benchmark Design}\label{sec:design}
The SupermarQ quantum benchmark suite is built around four guiding principles that shape the selection and evaluation of the applications. We start by motivating the design principles and then define the hardware-agnostic features used to characterize the quantum programs.

\subsection{Design Principles}\label{subsec:principles}
\textit{(1) Scalability} -- The current trajectory of QC development begins with the small-scale NISQ devices being built today and is aimed at the large-scale FT quantum computers of tomorrow. Because of this large variation in system size the applications included in a quantum benchmark suite should be gracefully scalable from just a few qubits to hundreds, thousands, and beyond -- while maintaining their meaning. For example, combinatorial optimization problems like MaxCut are scalable in this context because they can be defined on graphs of arbitrary size. It is also important that the performance metrics scale efficiently. Classical simulations of quantum circuits scale exponentially with the number of qubits so simply simulating the benchmarks and comparing with the experimental results is not a scalable solution. Therefore, a scalable suite must be composed of applications whose size is parameterizable and performance is easily verifiable.

\textit{(2) Meaningful and Diverse} -- Benchmark applications should reflect the workloads that will appear in practice. Potential use-cases for QPUs have been identified in chemistry~\cite{yung2014transistor, peruzzo2014variational}, machine learning~\cite{harrow2009quantum, biamonte2017quantum}, cryptography~\cite{shor1999polynomial, anschuetz2019variational}, finance~\cite{woerner2019quantum, braine2021quantum}, physics~\cite{low2019hamiltonian, uvarov2020machine}, and database search~\cite{grover1996fast}. Incorporating applications from a range of domains will provide relevant performance points to the widest range of people. Quantum programs pulled from different use-cases present wildly varying program structures and require different amounts of resources from the quantum computer. A benchmark suite should provide good coverage over these potential use-cases to better understand system performance under a variety of circumstances. The feature vectors introduced in Sec.~\ref{subsec:features} are a step in quantifying the stress an application places on a QPU.

\textit{(3) Full-system evaluation} -- The overall performance of a quantum computer relies on the proper functioning and interplay between the hardware and software stacks. Within the current stage of QC, the role played by the compiler: effectively cancelling gates, mapping between program and physical qubits, and so on, can make or break the execution of a quantum program~\cite{murali2019noise, resch2021error}. In addition, many of the unique properties offered by different quantum implementations (native multi-qubit or parameterizable gates for example) are exploited at the compiler level when the program is transpiled to a hardware supported gateset. 

Mandating a single compilation toolflow is inefficient, requiring that each benchmark be represented as an executable for every hardware backend, and ineffective, since certain capabilities available only to a certain class of quantum hardware may be overlooked. An application-based quantum benchmark suite should therefore specify benchmarks at a shared level of abstraction, such as OpenQASM, and allow the compiler to play a role in overall system performance. 

\textit{(4) Adaptivity} -- The entirety of quantum computing, encompassing both the hardware and software, is undergoing a period of rapid advancement. This poses a challenge for benchmarking since any suite which aims to accurately measure performance must keep pace with the development of algorithms, compilation optimizations, and hardware. The applications making up the benchmark suite should reflect this by adapting to the current state-of-the-art.

\subsection{Feature Vectors}\label{subsec:features}
We use a set of feature vectors to quantify the coverage of the selected benchmark applications. The features indicate how each of the benchmarks will stress the processor and to what degree. 

\subsubsection{Program Communication}
Quantum algorithms vary in the amount of communication needed between qubits. Some algorithms only require single qubit operations and nearest-neighbor interactions. These algorithms are easily mapped to processors with limited connectivity between qubits. Other algorithms require communication between every pair of qubits. Within a quantum circuit, a qubit's ``degree'' is the number of other qubits it interacts with via multi-qubit operations. Node degree is commonly used for physical architecture analysis in classical~\cite{jantsch2003networks} and quantum networking~\cite{bapat2018unitary}. It is often the case that physical qubit degree is much more uniform and limited than what is required for logical algorithm qubits. For hardware with less than all-to-all connectivity, the compiler may need to insert swap operations into the program to successfully map between the algorithmic and physical qubits~\cite{li2019tackling}. We use the normalized average degree of the program's interaction graph to quantify the communication requirements of quantum circuits. The interaction graph is formed by taking the qubits to be the vertices and inserting an edge between every pair qubits that interact with one another. The program communication feature is computed by taking the average degree of the interaction graph divided by the average degree of a complete graph with an equivalent number of qubits.
The program communication feature is computed as
\begin{equation}
    \mathit{C} = \frac{\sum_i^N d(q_i)}{N(N-1)}
    \label{eqn:con}
\end{equation}
for an $N$-qubit circuit, where $d(q_i)$ is the degree of qubit $q_i$. The communication requirements of sparsely connected applications will have values near zero while denser programs will be close to one.

\subsubsection{Critical-Depth}
The lifetime of the information stored across a QPU's qubits, the coherence time, is limited. This limitation combined with accumulated gate error causes lower fidelity circuit executions. Thus, it is essential that quantum circuits are of the shortest duration possible. The minimum duration for a quantum circuit is determined by the critical path: the longest span of dependent operations from circuit input to output. The critical path is a valuable benchmarking metric because quantum hardware performance must reach specific thresholds to accommodate continuously compounding gate errors. Operations of particular interest are two-qubit interactions because two-qubit operations dominate single-qubit operations in terms of gate error and execution time on NISQ hardware~\cite{pino2021demonstration}~\cite{jurcevic2021demonstration}. The critical-depth feature gives context about how many two-qubit interactions in a program lie along the critical path and contribute to the overall circuit depth. It is calculated as
\begin{equation}
    \mathit{D} = n_{e_d} / n_e
\end{equation}
where $n_{e_d}$ is the number of two-qubit interactions on the longest path that sets the circuit depth and $n_{e}$ is the total number of two-qubit interactions in the circuit. Circuits that are heavily serialized will have a critical-depth that's close to 1. 

\subsubsection{Entanglement-Ratio}
Entanglement is a critical property which gives quantum computing much of its strength. It makes for a useful benchmark for quantum machine performance as it can be applied to computing tasks that demonstrate quantum advantage such as in Shor's factoring~\cite{shor1999polynomial}, teleportation~\cite{bennett1993teleporting}, superdense coding~\cite{bennett1992communication}, and quantum cryptographic protocols~\cite{ekert1991quantum}. Prior work indicates that algorithms without entanglement can be efficiently simulated by classical computers~\cite{vidal2003efficient,verstraete2004matrix}; further demonstrating the importance of entanglement as a benchmark for quantum processing power. While it is in general quite difficult to measure the precise amount of entanglement at every point within a circuit (usually requiring access to the full statevector) we can roughly capture this feature by computing the proportion of all gate operations ($n_g$) which are two-qubit interactions ($n_e$): 
\begin{equation}
    \mathit{E} = n_e / n_g.
\end{equation}

\subsubsection{Parallelism}
The structure of different quantum algorithms allow for varying degrees of parallelization. Parallel operations can also stress the quantum hardware because of correlated noise events known as ``cross-talk'' that degrade program performance~\cite{murali2021instruction}. Cross-talk, often caused by simultaneous gate execution, is a common source of error in NISQ systems, and its negative impact on program execution has been well studied~\cite{murali2020software,ding2020systematic}. This motivates the development of a feature that captures how susceptible a benchmark is to degradation via cross-talk. The parallelism feature represents this aspect by comparing the ratios of the number of qubits ($n$), gates ($n_g$), and the circuit depth, $d$: 
\begin{equation}
    \mathit{P} = \left( \frac{n_g}{d}-1 \right) \frac{1}{n-1}.
\end{equation}
Highly parallel applications fit a large number of operations into a relatively small circuit depth and will therefore have a parallelism feature close to 1.

\subsubsection{Liveness}
During program execution, a qubit will either be involved in computation or it will be idle; waiting for its next instruction. In an ideal environment, the qubit's state would stay coherent while idling. In reality, unwanted environmental interactions such as amplitude damping, dephasing, and correlated noise cause decoherence~\cite{viola1999dynamical}. 
The liveness feature captures aspects of an application's qubit status during its lifetime. It can be defined as 
\begin{equation}
    \mathit{L} = \frac{\sum_{ij}A_{ij}}{n d},
\end{equation}
where $A$ is the liveness matrix defined by taking a quantum circuit and forming a matrix with $n$ rows equal to the number of qubits and a number of columns equal to the circuit depth $d$. At every time-step of circuit execution (i.e., each column), a qubit may either be involved in an operation or idle, corresponding to entries of 1 or 0 in the liveness matrix, respectively. In this way, the liveness feature gives a sense of how often the qubits are being acted upon. The frequency of idling as $1-L$ provides insight to qubit inactivity over its application lifetime.

\subsubsection{Measurement}
Qubit-specific measurement is a critical part of quantum computing~\cite{divincenzo2000physical}. It is required to extract information during and after a program's execution. In fault-tolerant quantum computing, error correcting codes use measurement to extract entropy from a noisy quantum system~\cite{preskill1998fault}. Unfortunately, NISQ devices suffer from non-trivial amounts of measurement error. The measurement feature, 
\begin{equation}
    \mathit{M} = l_{mcm} / d
\end{equation}
focuses specifically on the mid-circuit measurement and reset operations within a quantum program. For a circuit composed of $d$ sequential layers of gate operations (i.e.,~the circuit depth), $l_{mcm}$ is the number of layers which contain these measurement and reset operations.

\begin{figure*}[t!]
     \centering
     \begin{subfigure}[b]{\textwidth}
         \centering
         \includegraphics[width=\textwidth]{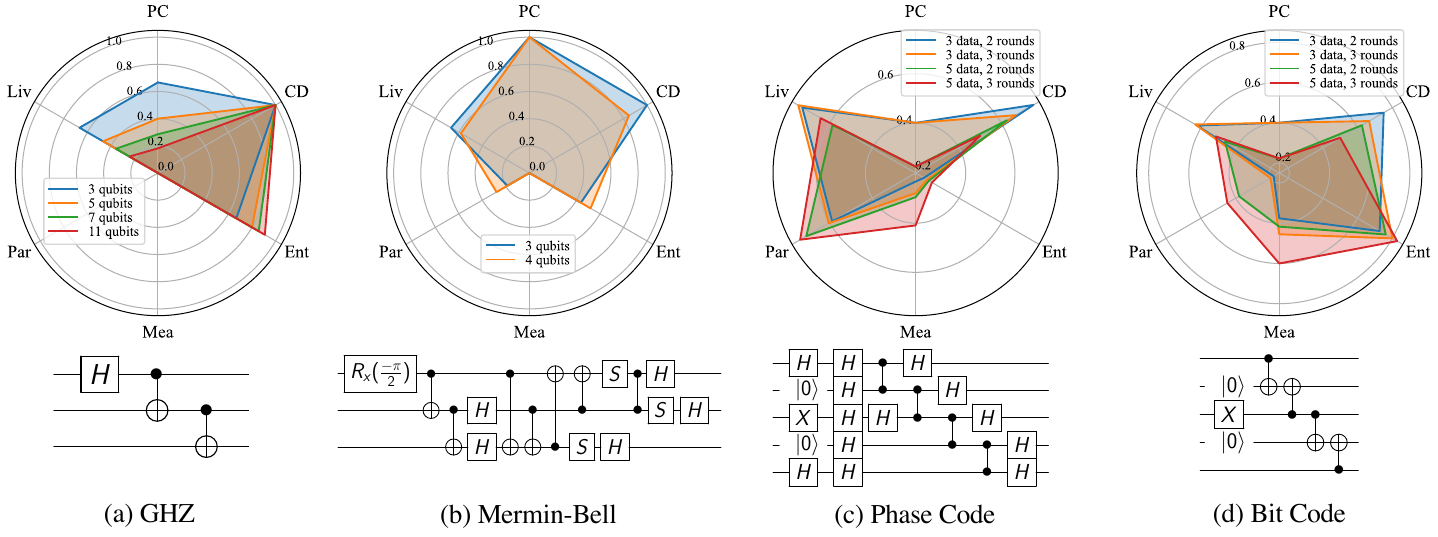}
     \end{subfigure}
     
     \medskip
     \begin{subfigure}[b]{\textwidth}
         \centering
         \includegraphics[width=\textwidth]{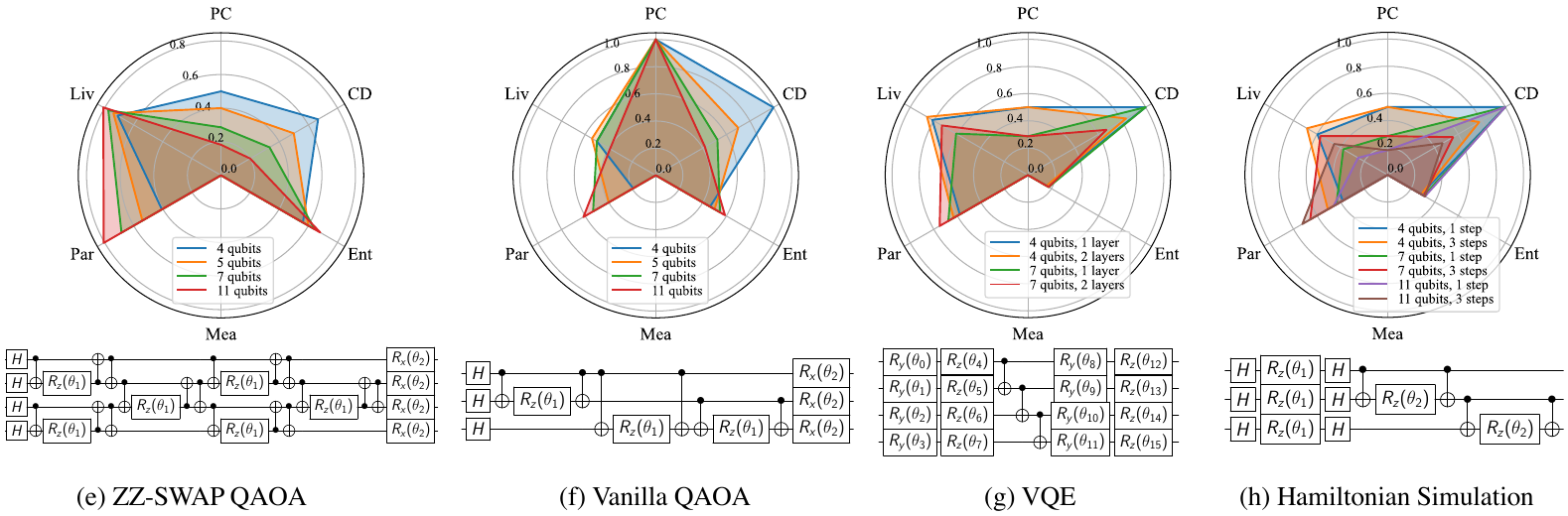}
     \end{subfigure}
    \caption{Feature maps and sample circuits for each of the benchmarks evaluated in this study. The definitions of the Program Communication (PC), Critical Depth (CD), Entanglement-Ratio (Ent), Measurement (Mea), Parallelism (Par), and Liveness (Liv) features are given in Sec.~\ref{sec:design}.}
    \label{fig:features}
\end{figure*}

\section{Benchmark Applications}\label{sec:benchmarks}

\subsection{GHZ}
The generation of entanglement between qubits is one of the most important tasks in quantum computing, sensing, and networking. We benchmark the ability of a quantum processor to generate entanglement by measuring the state preparation fidelity of GHZ states~\cite{greenberger1989bell}. The GHZ benchmark consists of a Hadamard gate followed by a ladder of CNOTs to produce the entangled state: $(\ket{00 \dots 0} + \ket{11 \dots 1}) / \sqrt{2}$ (see Fig.~\ref{fig:features}a). The performance metric is the Hellinger fidelity~\cite{harper2020efficient, qiskit2021hellinger} between the experimentally observed probability distribution and the ideal distribution ( $50\%$ $\ket{00 \dots 0}$ and $50\%$ $\ket{11 \dots 1}$).

There are other methods for preparing GHZ states, notably those utilizing mid-circuit measurements or parallel two-qubit gates. These methods can have different resource requirements in terms of gate counts and circuit depth~\cite{nation2021johnson, mooney2021generation}. However, we choose to include the CNOT-ladder method because not all platforms currently support mid-circuit measurements.

\subsection{Mermin-Bell}
One of the primary uses for quantum computers thus far has been for small scale demonstrations of the quantumness of nature~\cite{mooney2019entanglement, garcia2017five}. These experiments are known as Bell inequality tests~\cite{bell1964einstein} whose introduction resolved the Einstein-Podolsky-Rosen (EPR) paradox that questioned the completeness of quantum mechanics~\cite{einstein1935can}. The Mermin-Bell benchmark (Fig.~\ref{fig:features}b) included in SupermarQ is an example of a Bell inequality test. In this benchmark, a GHZ state, $\ket{\phi} = (1/\sqrt{2})(\ket{00\dots0} + i\ket{11\dots1})$, is first prepared before measuring the expectation value of the Mermin operator
\begin{equation}\label{eqn:mermin}
    M = \frac{1}{2i} \left( \prod_{j=1}^n (\sigma^j_x + i \sigma^j_y) - \prod_{j=1}^n (\sigma^j_x - i \sigma^j_y) \right)
\end{equation}
where $\sigma^j_x$ and $\sigma^j_y$ are the Pauli-X and -Y operators acting on the $j$-th qubit.
If nature is quantum, the expectation of this operator for an $n$ qubit system is
\begin{equation}
    \expval{M}{\phi} = 2^{n-1}.
\end{equation}
If nature is classical and obeys a theory of local-hidden variables, then the expectation value of the Mermin operator is bounded by
\begin{equation}
\expval{M}{\phi} \leq 2^{\left(n - (n \text{ mod } 2) \right)/2}
\end{equation}
We measure performance by computing $(\expval{M}{\phi} + 2^{n-1}) / 2^n$ as the benchmark score.

After preparing the GHZ state, the remaining gates within the Mermin-Bell circuits rotate the quantum state into the shared basis of the Mermin operator such that the expectation of each term can be measured simultaneously. Unlike the GHZ benchmark, the basis-change portion of the circuit begins to dominate the state preparation as the size of the benchmark increases. 

\subsection{Error Correction Subroutines}
Error correcting codes (ECCs) are the means by which fault-tolerant quantum computers are able to execute arbitrarily long programs. Many ECCs have been developed that trade off between the number of detectable errors, correctable errors, qubits required, and required error thresholds to reach fault-tolerance~\cite{divincenzo1996fault, calderbank1998quantum, fowler2012surface}. Although full-scale fault-tolerance has not yet been observed, small experiments have demonstrated the feasibility of different error correction schemes on both superconducting and trapped ion architectures~\cite{egan2020fault, chen2021exponential, ryan2021realization}.

Since the error levels of current NISQ devices do not allow for the implementation of full-scale error correction, we use two proxy-applications to benchmark QPU performance within this domain. While these proxy-applications do not correct any errors, they do reflect the circuit structure that is common to many ECCs~\cite{preskill1998fault, reed2012realization}. Unlike the other benchmarks within the SupermarQ suite, the error correction proxy-applications make use of \textsc{reset} operations (needed to reinitialize a qubit to the $\ket{0}$ state after measurement). The data qubits which do not participate in the \textsc{reset} will need to idle. This idleness will add to the circuit execution time; increasing the chances of decoherence.

\subsubsection{Phase Code Proxy-application}
The phase code benchmark is a phase flip repetition code parameterized by the number of data qubits and rounds of error correction. The feature maps for different parameterizations are shown in Fig.~\ref{fig:features}c as well as a sample circuit which has three data qubits and a single round of error correction. To measure performance, we first prepare the data qubits in initial $\ket{+} = (\ket{0}+\ket{1})/\sqrt{2}$ or $\ket{-} = (\ket{0}-\ket{1})/\sqrt{2}$ states followed by $r$ rounds of error correction and finally a measurement of the final state. In a noiseless setting, the final state of the system is known a priori: it should be identical to the chosen initial state. We therefore compute the Hellinger fidelity between the experimental and ideal distributions as a measure of performance. For example, the data qubits in Fig.~\ref{fig:features}c's sample circuit are initialized in the $\ket{+-+}$ state and the ideal output distribution is an equal distribution over all the possible values of the three data qubits and the error-syndrome qubits in the $\ket{00}$ state. 

\subsubsection{Bit Code Proxy-application}
Like the phase code, the bit code benchmark is also a bit flip repetition code that is parameterized by the number of data qubits and error correction rounds. Instead of checking for phase flips, the bit code detects bit flips on the data qubits. Fig.~\ref{fig:features}d shows the feature map for this benchmark and a sample circuit with three data qubits initialized in the $\ket{010}$ state and a single round of error correction. Since the ideal final state is known a priori, we also use the Hellinger fidelity as the score function for this benchmark.

\subsection{QAOA}
The Quantum Approximate Optimization Algorithm (QAOA) is a variational quantum-classical algorithm that can be trained to output bitstrings to solve combinatorial optimization problems~\cite{farhi2014quantum}. 
We benchmark QAOA for MaxCut on complete graphs with edge weights randomly drawn from $\{-1, +1\}$. This is known as the Sherrington-Kirkpatrick (SK) model; and it is a particularly promising target for near-term quantum computers \cite{farhi2019sherrington, harrigan2021quantum}. We implement two variants of QAOA that use different parameterized circuits (ansatzes). 

The Vanilla QAOA benchmark, Fig.~\ref{fig:features}f, uses an ansatz that matches the SK model exactly. This is the typical formulation of QAOA~\cite{farhi2014quantum}. Since the SK model is completely connected, the constructed ansatz also requires all-to-all connectivity. The ZZ-SWAP QAOA benchmark implements a variational ansatz known as a SWAP network~\cite{kivlichan2018quantum, tomesh2020coreset}. This ansatz is a natural choice for solving MaxCut on the SK model which requires an interaction between every pair of qubits (i.e., $n(n-1)/2$ edges). The SWAP network (a sample circuit is shown in Fig.~\ref{fig:features}e) is able to perform all $O(n^2)$ required interactions using a quantum circuit whose depth scales as $O(n)$.

We use a proxy-application in place of the full variational algorithm due to current limitations associated with cloud-based access to QC systems. The full QAOA benchmark would require thousands of iterations to reach convergence. Evaluating the full benchmark becomes infeasible because of the wait times incurred while the jobs are in the queue. We measure a QPU's ability to evaluate a single iteration of QAOA instead.

To ensure scalable classical verification, we choose the level-one ($p = 1$) variant of QAOA; which is efficiently simulable classically due to recent work \cite{wang2018quantum}.
We found optimal parameters via classical simulation and then executed these QAOA circuits on the real QC systems. We compared the experimental and ideal results by measuring the expectation value, $\langle H \rangle$, and computing $1 - \abs{\frac{\expval{H}_{ideal} - \expval{H}_{exper}}{2\expval{H}_{ideal}}}$ as the benchmark score. For the SK model, this can be written as $H = \sum_{i,j \in E}{\sigma_z^i \sigma_z^j}$;
where $E$ is the set of edges within the graph. In contrast, the performance measure for the full QAOA benchmark would be the final MaxCut value achieved after optimization. This would allow for straightforward comparisons with other quantum or classical MaxCut algorithms.

\subsection{VQE}
The Variational Quantum Eigensolver (VQE)~\cite{peruzzo2014variational} is another hybrid algorithm like QAOA. The goal of this algorithm is to find the lowest eigenvalue of a given problem matrix by computing a difficult cost function on the QPU and feeding this value into an optimization routine running on a CPU. Typically, the problem matrix is the Hamiltonian governing a target system and the lowest eigenvalue corresponds to the system's ground state energy~\cite{mcclean2016theory}.

We target the one dimensional transverse field Ising model (TFIM, also called the transverse Ising chain) and use VQE to find its ground state energy. The 1D TFIM is a useful model for understanding phase transitions in magnetic materials~\cite{uvarov2020machine}. The 1D TFIM is desirable as a scalable benchmark because it is exactly solvable via classical methods~\cite{pfeuty1970one}.

Like the proxy-application employed for the QAOA benchmark, we replace the full VQE benchmark with a proxy-application that measures performance for a single iteration of the VQE algorithm. Instead of running the full VQE algorithm and reporting the final ground state energy, we classically simulate the variational optimization to convergence. We take the final parameters output by said classical optimization and measure the energy of the 1D TFIM using the quantum computer. We compare this energy with the value obtained classically and compute the same score function as the QAOA benchmark. The hardware-efficient ansatz used in this benchmark is shown in Fig.~\ref{fig:features}g along with its corresponding feature map.

\subsection{Hamiltonian Simulation}
Simulating the time evolution of quantum systems is one of the most promising applications of quantum computing~\cite{lloyd1996universal}. There are many quantum algorithms for Hamiltonian simulation which are known to possess exponential speedups over classical methods~\cite{campbell2019random, low2019hamiltonian}. Closing the gap between the algorithmic resource requirements and the capabilities of QC systems may lead to breakthroughs in the development of new batteries and catalysts~\cite{reiher2017elucidating}.

We target the 1D TFIM as the system we wish to simulate.
The Hamiltonian for this system, consisting of $N$ spins, may be written as
\begin{equation}
    H = - \sum_{i=1}^N (J_z \sigma_z^i \sigma_z^{i+1} + \epsilon_{ph} \cos{(\omega_{ph} t)} \sigma_x^i)
    \label{eqn:ising hamiltonian}
\end{equation}
where $J_z$ is a coupling constant that determines the strength of the nearest-neighbor interactions and $\epsilon_{ph}\cos{\omega_{ph} t}$ describes the time-varying magnetic field. We set these parameters to match recent work on quantum algorithms for simulating the time evolution of quantum systems~\cite{bassman2020towards}.

The Hamiltonian simulation benchmark (Fig.~\ref{fig:features}h) is specified by taking the Hamiltonian in Eq.~\ref{eqn:ising hamiltonian} for a specific value of $N$, generating a quantum circuit via Trotterization~\cite{yung2014introduction} for a specific number of time steps, and finally measuring the average magnetization of the final state. The average magnetization of the final quantum state can be found by computing the expectation value of the operator $m_z = \frac{1}{N}\sum_i{\sigma_z^i}$~\cite{bassman2020towards}. The experimentally obtained average magnetization is then compared to the exact value obtained classically. We compute $1 - \frac{\abs{\expval{m_z}_{ideal} - \expval{m_z}_{exper}}}{2}$ as the benchmark score.

\begin{table}[]
\centering
\begin{tabular}{c|c|c}
Suite     & Volume  & Circuits \\ \hline
SupermarQ (this work) & 9.0e-03 & 52       \\
QASMBench~\cite{li2020qasmbench} & 4.0e-03 & 62       \\
Synthetic & 1.4e-03 & 6 \\
CBG2021~\cite{cornelissen2021scalable}    & 1.6e-08 & 10476    \\
TriQ~\cite{murali2019full}      & 4.1e-14 & 12 \\
PPL+2020~\cite{patel2020experimental}    & 1.0e-15 & 9
\end{tabular}
\caption{Coverage comparison of different benchmark suites. For each suite we report the volume and the number of circuits used to compute the volume.}
\label{tab:coverage}
\end{table}

\subsection{Coverage}\label{subsec:coverage}
To analyze suite coverage we consider the volume of feature space spanned by the benchmarks. We treat the six application features as separate axes within a six dimensional space. Each benchmark within a suite can be associated with a single, six dimensional feature vector. To find the coverage of a given set of applications, we compute the volume of the convex hull defined by their feature vectors: each shape in the feature maps (Fig.~\ref{fig:features}) shown above corresponds to a single vector within the higher dimensional feature space. 

\begin{table*}[t!]
    \centering
    \includegraphics[width=\textwidth]{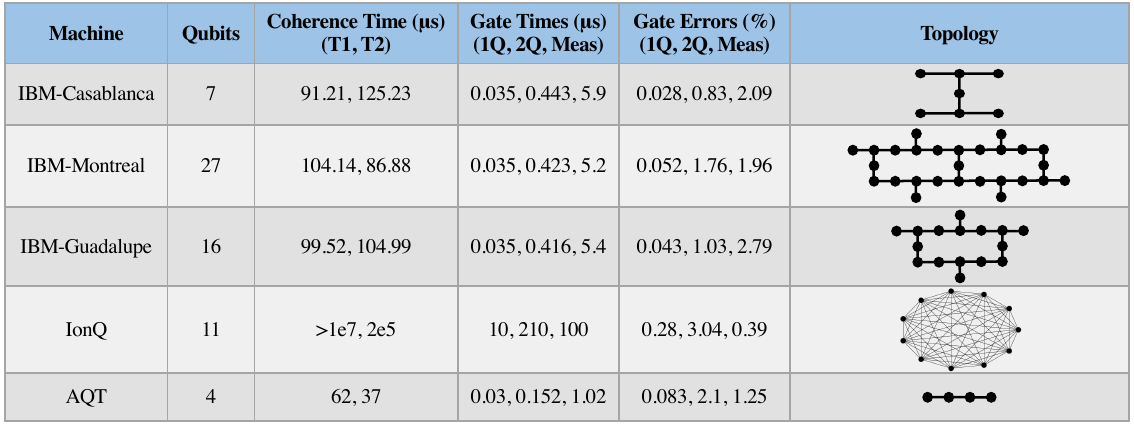}
    \caption{Characteristics of the QC systems used to evaluate the benchmarks. The IBM and IonQ data was taken from the public documentation available through their respective cloud providers (IBM Qiskit and AWS Braket) on July 30, 2021. The device statistics for the IBM QPUs not pictured here are available online through IBM Quantum~\cite{ibmquantum}. The AQT system properties were obtained via randomized benchmarking on Sept 21, 2021 and~\cite{hashim2020randomized}.
    }
    \label{tab:hardware table}
\end{table*}

We compute the coverage of six different quantum benchmark suites and report their volumes and the number of circuits used to compute the coverage in Table~\ref{tab:coverage}. QASMBench is a collection of benchmark circuits that range in size from two to a thousand qubits~\cite{li2020qasmbench}. CBG2021 is a recent suite that includes six different benchmark applications that range from Mermin-Bell tests to calculations of the Mandelbrot set~\cite{cornelissen2021scalable}. The TriQ suite was used in recent cross-platform comparisons between superconducting and trapped ion processors, and consists of small-scale applications with no more than eight qubits~\cite{murali2019full}. PPL+2020 introduced the ``quality of operation'' metric to capture the fidelity and variance of quantum gate operations, and is composed of nine small benchmark applications with three to five qubits~\cite{patel2020experimental}. For the SupermarQ suite, we generated instances of the applications covered in Sec.~\ref{sec:benchmarks} ranging in size from three to a thousand qubits. Finally, the synthetic suite consists of a set of hypothetical proxy-benchmarks that each maximize a single application feature (e.g., unit vectors along each axis of the six dimensional space).

Only SupermarQ and QASMBench attain coverage superior to the synthetic benchmark suite. These are also the only suites that include larger applications relevant to late NISQ and early FT devices. For comparison, the SupermarQ applications used in this coverage computation were selected to match the range of benchmark sizes found in the QASMBench suite, however, SupermarQ has the additional capability of generating arbitrarily sized benchmarks.

Periodically collecting new benchmark data is a practical concern for any quantum benchmark suite. We utilize a write-once-target-all toolflow, SuperstaQ, which was designed explicitly with this purpose in mind~\cite{superstaq}. With SuperstaQ we are able to specify the OpenQASM for a single circuit and execute it on multiple backends. The need to efficiently collect new benchmark results also introduces a tradeoff between the number of circuits in the suite (more circuits covering more applications and boosting coverage) and the ability to evaluate them in a cost-efficient manner. SupermarQ tries to find a balance between the two; providing competitive coverage that is superior to a purely synthetic suite while using a relatively modest number of circuits.

\section{Methodology}\label{sec:methodology}

\begin{figure*}[t!]
     \centering
     \begin{subfigure}[b]{0.49\textwidth}
         \centering
         \includegraphics[width=\linewidth]{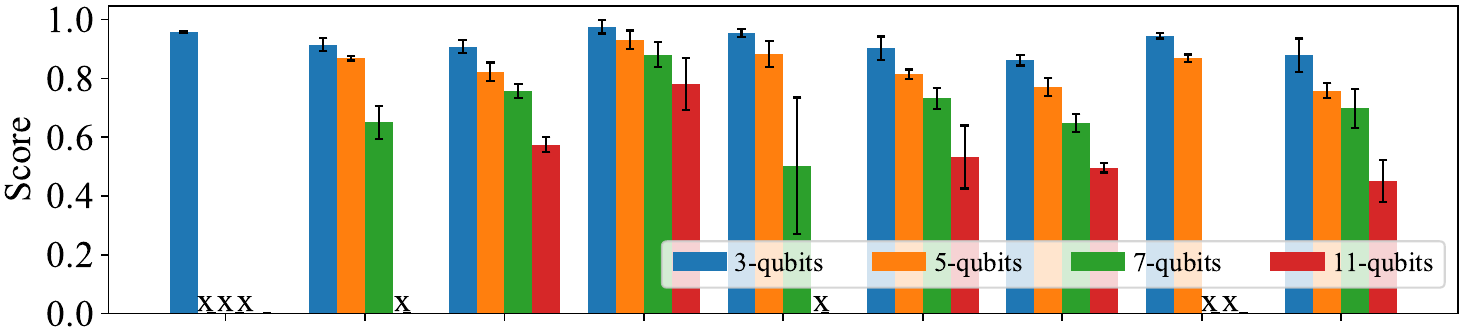}
         \caption{GHZ}
         \label{fig:ghz-scores}
     \end{subfigure}
     \hfill
     \begin{subfigure}[b]{0.49\textwidth}
         \centering
         \includegraphics[width=\linewidth]{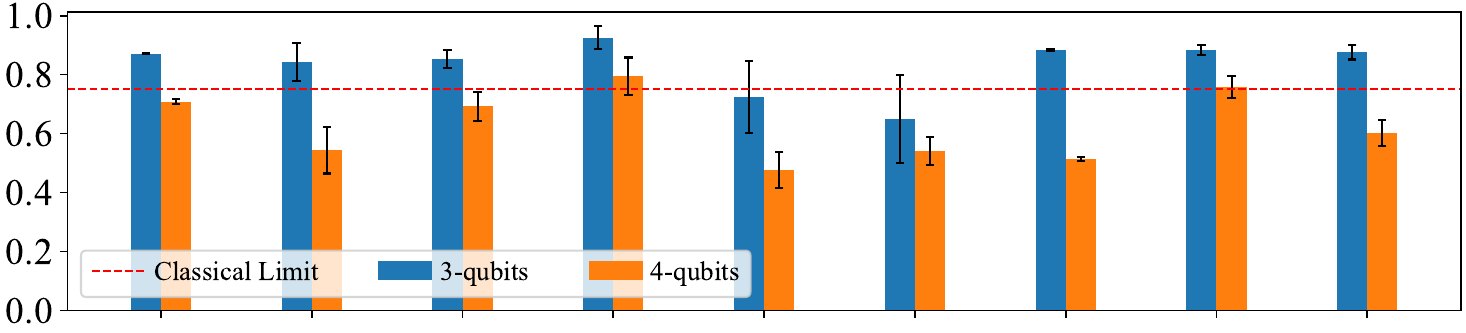}
         \caption{Mermin-Bell}
         \label{fig:merminbell-scores}
     \end{subfigure}
     
     \medskip

     \begin{subfigure}[b]{0.49\textwidth}
         \centering
         \includegraphics[width=\linewidth]{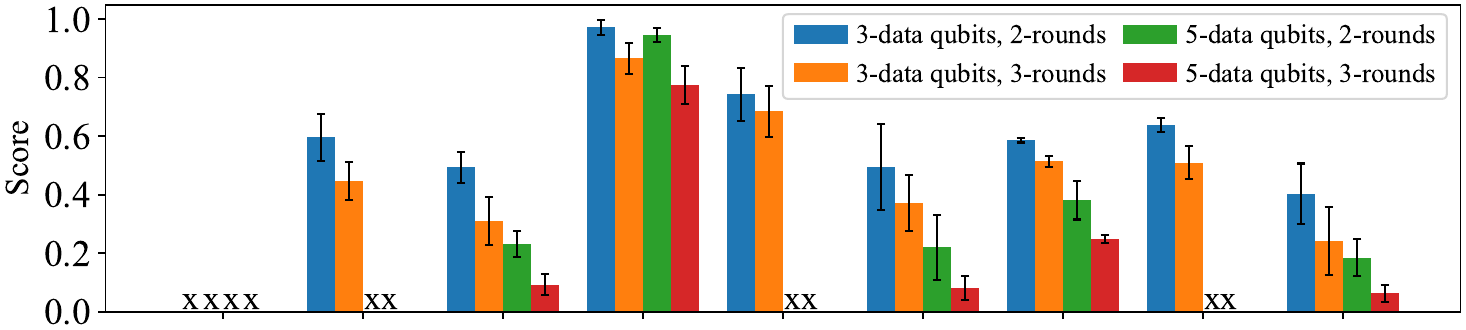}
         \caption{Bit Code}
         \label{fig:bitcode-scores}
     \end{subfigure}
     \hfill
     \begin{subfigure}[b]{0.49\textwidth}
         \centering
         \includegraphics[width=\textwidth]{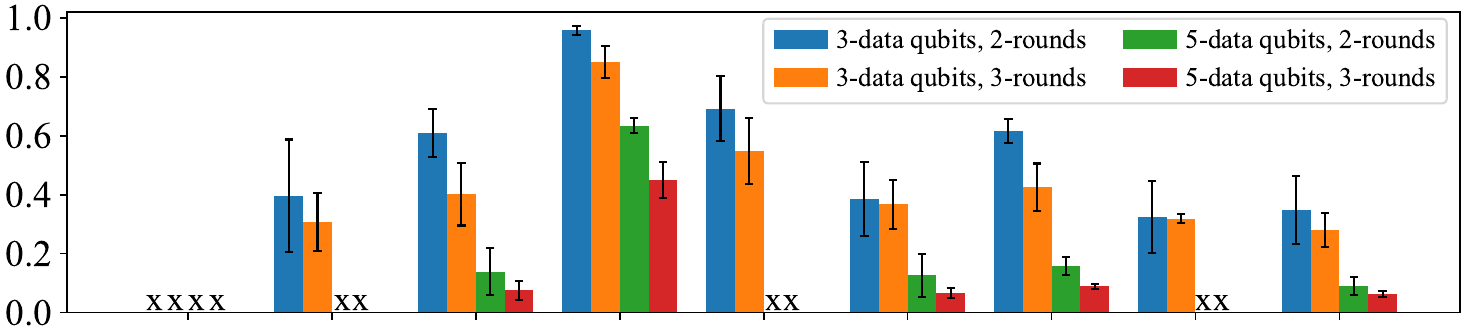}
         \caption{Phase Code}
         \label{fig:phasecode-scores}
     \end{subfigure}
     
     \medskip
     
     \begin{subfigure}[b]{0.49\textwidth}
         \centering
         \includegraphics[width=\linewidth]{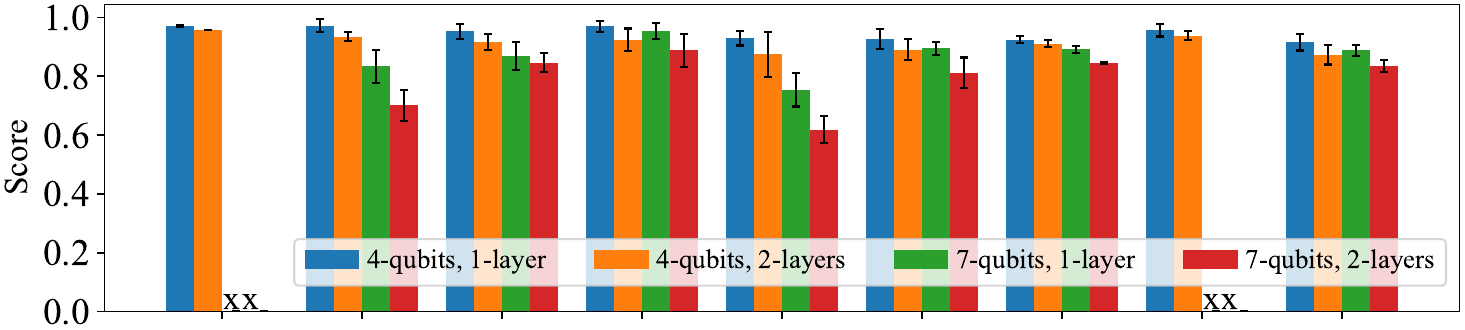}
         \caption{VQE}
         \label{fig:vqe-scores}
     \end{subfigure}
    \hfill
     \begin{subfigure}[b]{0.49\textwidth}
         \centering
         \includegraphics[width=\linewidth]{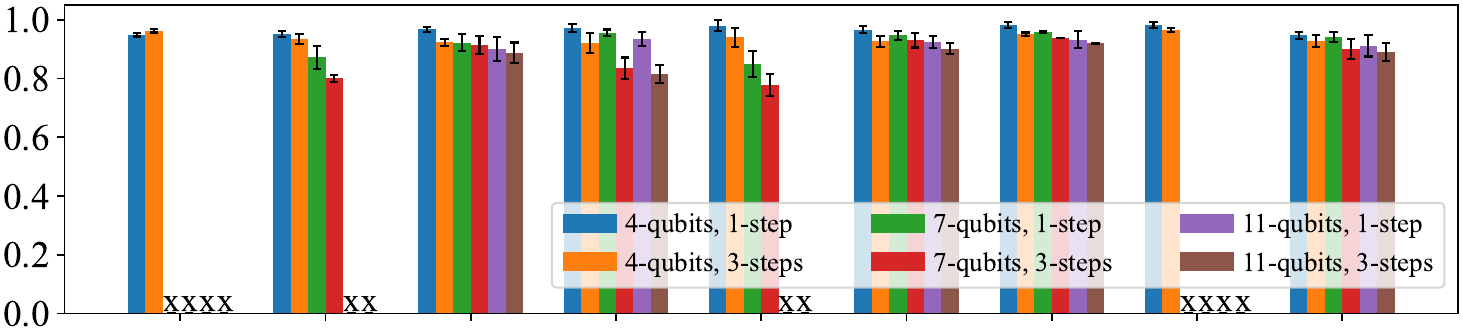}
         \caption{Hamiltonian Simulation}
         \label{fig:hs-scores}
     \end{subfigure}
     
     \medskip
     
     \begin{subfigure}[b]{0.49\textwidth}
         \centering
         \includegraphics[width=\linewidth]{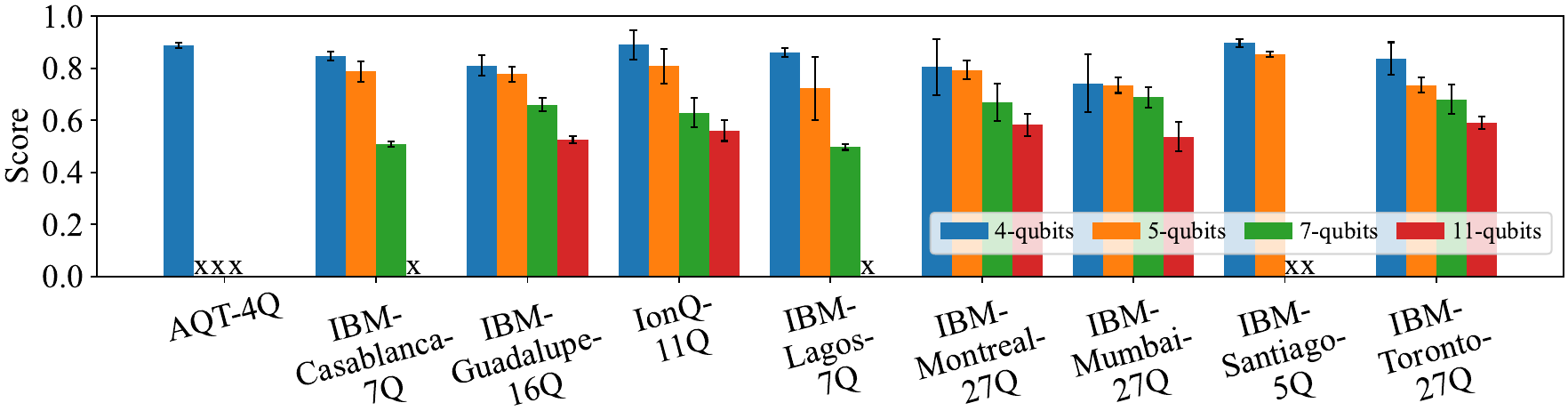}
         \caption{ZZ-SWAP QAOA}
         \label{fig:qaoafswap-scores}
     \end{subfigure}
     \hfill
     \begin{subfigure}[b]{0.49\textwidth}
         \centering
         \includegraphics[width=\linewidth]{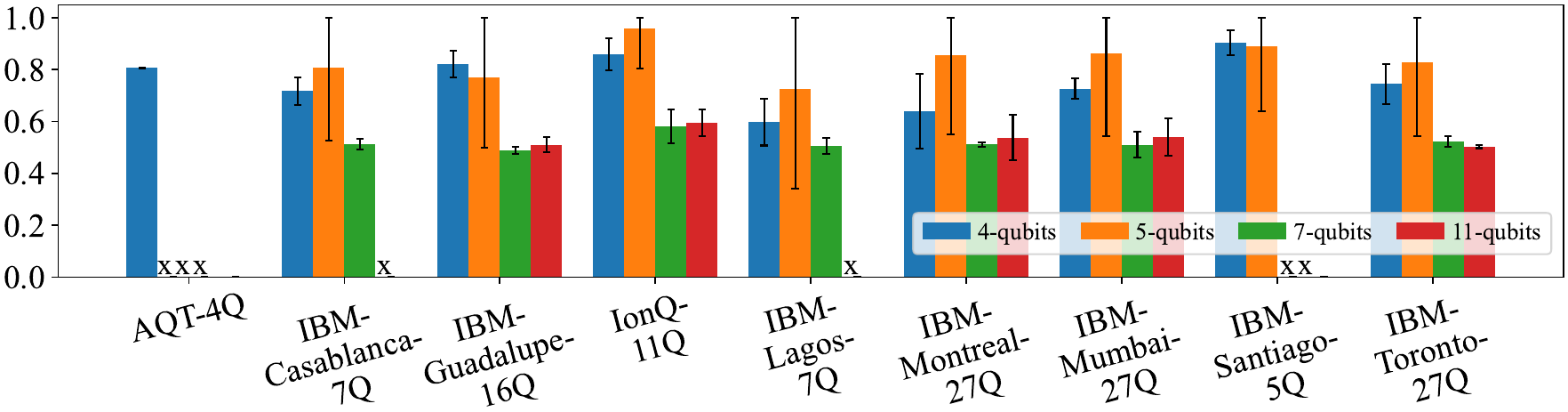}
         \caption{Vanilla QAOA}
         \label{fig:qaoavanilla-scores}
     \end{subfigure}
    \caption{Benchmark results evaluated across superconducting and trapped ion devices (the black X's indicate benchmarks that exceed the number of qubits available on the device). The results for each benchmark appear in the same order given along the x-axis of (g) and (h). Each bar denotes the average performance over multiple benchmark runs while the error-bars indicate a single standard deviation from the mean score. The specific score functions for each benchmark are given in Sec.~\ref{sec:benchmarks}. In every benchmark run, we executed 2000 shots on the IBM devices, 1024 on the AQT device, and 35 on the IonQ processor. The shot counts were selected to maintain a reasonable cost budget for collecting the benchmark results.}
    \label{fig:results}
\end{figure*}

In this work we present results obtained for eight benchmark applications evaluated on nine QPUs. We accessed the quantum computers through the IBM Qiskit~\cite{ibmquantum} and AWS Braket~\cite{braket} cloud services, and the Lawrence Berkeley National Lab's Advanced Quantum Testbed (AQT)~\cite{hashim2020randomized}. 
The specifics of each benchmark's evaluation and score function are given in Sec.~\ref{sec:benchmarks}, and the architectural characteristics of the quantum computers used to evaluate the suite of benchmarks are summarized in Table~\ref{tab:hardware table}.

For each benchmark we first fix the application-specific parameters (e.g., problem size, number of layers, initial state). Then the OpenQASM for the benchmark circuits is generated. Some benchmarks may be composed of multiple circuits. For example, the VQE benchmark requires two separate circuits in order to measure the energy operator in two orthogonal bases.

To easily evaluate the benchmarks across QPUs we utilize SuperstaQ~\cite{superstaq}: a write-once-target-all toolflow which presents a unified interface for simultaneously submitting OpenQASM-defined quantum circuit instances to the devices available on the IBM Qiskit and AWS Braket cloud services. Behind the scenes, SuperstaQ converts OpenQASM to AWS Braket's jaqcd (JsonAwsQuantumCircuitDescription) intermediate representation \cite{braket_schemas}. In addition to thorough unit tests and unitary-verification integration tests, we experimentally validated the correctness of SuperstaQ by running our error correction benchmarks for a comprehensive set of input-output bitstring pairs. IBM's Qiskit supports OpenQASM out-of-the-box, so it does not require any conversion.

Part of the challenge associated with evaluating the benchmarks in this suite stems from the fact that the level of control over which compiler optimizations are applied to the circuits varies across the different cloud services. 
SupermarQ enables cross-platform comparisons of performance by specifying its benchmarks at a shared level of abstraction. To do this, we evaluate all the applications within the context of a \textit{Closed Division}, that specifies how the benchmarks are expressed and the optimizations that are allowed.

The Closed Division allows for a restricted set of optimizations to obtain a lower bound for the performance of a quantum computer. The benchmarks in this suite are specified at the level of OpenQASM~\cite{cross2017open}, the most popular \cite{singhal2020verified} intermediate representation for quantum circuits. Optimizations which are publicly available to quantum programmers are considered fair-game. These include the transpilation of OpenQASM to native gates, noise-aware qubit mapping, SWAP insertions, reordering of commuting gates, and cancellation of adjacent gates. Low-level optimizations below the level of native gates, such as pulse optimizations, as well as post-processing techniques like error-mitigation are not allowed. The optimizations included within the Closed Division were chosen to match the optimizations that are automatically applied when using the cloud-based platforms. This matches the level of optimization that would be available to the average user.

The specification of the Closed Division and the benchmark results presented in this work aim to demonstrate a lower bound on the performance which would be achievable by a typical quantum programmer. We leave the specification and evaluation of an Open benchmarking division, allowing for a wider range of optimizations, for future work. The goals of these two benchmarking divisions parallel the design of the MLPerf benchmark suite~\cite{mattson2020mlperf}.

\section{Results}\label{sec:results}

\begin{figure*}[t]
    \centering
    \begin{subfigure}{.5\textwidth}
      \centering
      \includegraphics[width=\linewidth]{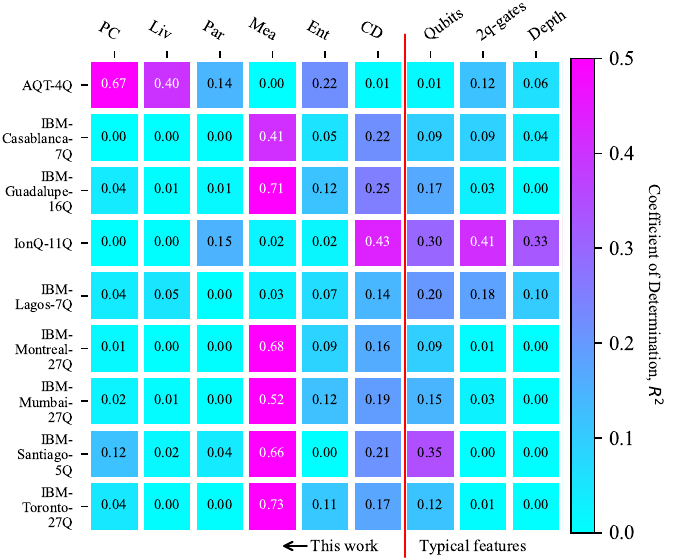}
      \caption{Including error-correction benchmarks.}
      \label{fig:heatmap}
    \end{subfigure}%
    \begin{subfigure}{.5\textwidth}
      \centering
      \includegraphics[width=\linewidth]{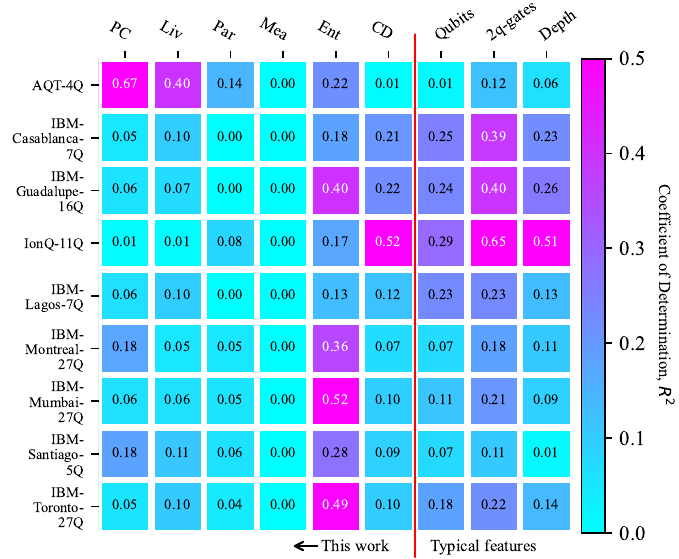}
      \caption{Excluding error-correction benchmarks.}
      \label{fig:heatmap_noEC}
    \end{subfigure}
    \caption{Heatmaps showing the correlation between application features and system performance. The correlations in (a) were computed using all of the benchmark data, whereas in (b) the data from the phase and bit code benchmarks was excluded.}
    \label{fig:correlations}
\end{figure*}

\begin{figure}
    \centering
    \includegraphics[width=\columnwidth]{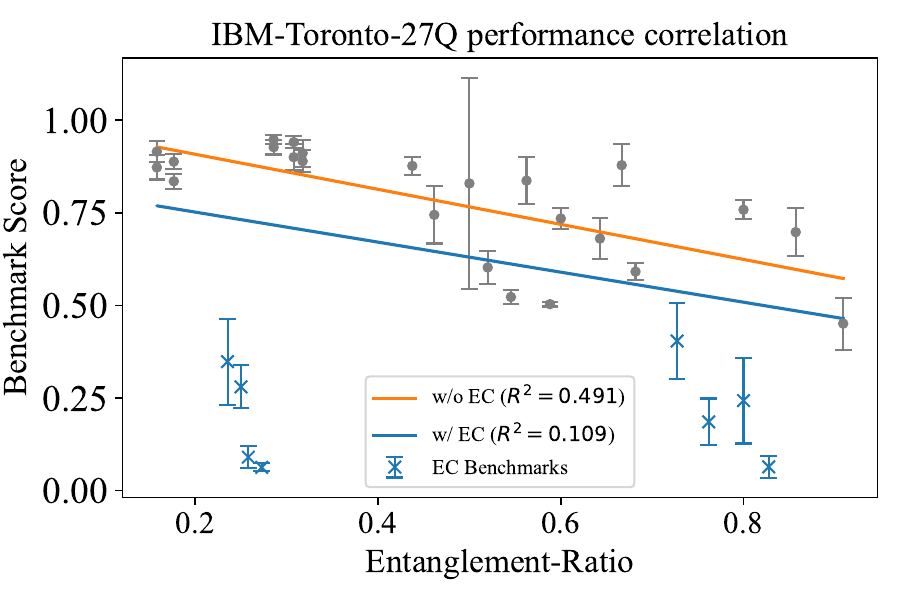}
    \caption{Example of the impact the error correction (EC) benchmarks have on the correlation between the application features and system performance.}
    \label{fig:toronto_correlation}
\end{figure}

The results of the benchmark executions are shown in Fig.~\ref{fig:results}. Benchmarks labeled with black X's were too large to fit on a device. As the width and depth of the benchmarks increases, the scores obtained by the hardware tends to decrease. This is expected as it is harder to maintain a coherent quantum state as the number of qubits and gate operations grows. There are also cases where adding additional qubits is less detrimental to performance than adding more gates. We see this behavior in the results of the bit code (IonQ), VQE (IonQ, Montreal, and Mumbai), and Hamiltonian simulation (IonQ, Mumbai, and Toronto) benchmarks.

The Mermin-Bell results shown in Fig.~\ref{fig:merminbell-scores} indicate that the QPUs are able to exploit quantum effects and surpass the classical limit denoted by the red line. However, this is still a difficult benchmark --- few processors are able to meet the classical limit for the 4-qubit instance. The high communication feature of the Mermin-Bell benchmark (Fig.~\ref{fig:features}b) reflects the all-to-all circuit structure necessary to measure the Mermin operator (Eq.~\ref{eqn:mermin}). Indeed, we see that the IonQ trapped ion device, which natively supports all-to-all connectivity, achieves the best performance despite having a higher two-qubit gate error rate than many of the superconducting devices.

The importance of compatibility between circuit structure and qubit topology is seen throughout the benchmark suite. Although many of the superconducting devices have two-qubit error rates lower than that of the trapped ion device, the additional swap operations that must be inserted to match the program connectivity quickly deteriorate performance (Mermin-Bell, vanilla QAOA). When the connectivity of the program matches that of the hardware, then the high quality gates of the superconducting QPUs results in competitive performance with the all-to-all connectivity of the trapped ion QPU (VQE, Hamiltonian simulation, ZZ-SWAP QAOA).

In Fig.~\ref{fig:heatmap} we show the correlations between the application features introduced in Sec.~\ref{sec:design} and the benchmark scores. For comparison, we also include typical features such as circuit depth and the number of qubits and two-qubit gates which have been used to characterize quantum applications in prior work~\cite{murali2019noise, murali2019full}. The coefficient of determination ($R^2$) for each feature-QPU pair can be interpreted as the proportion of the variance in that QPU's performance that is attributable to that feature. The $R^2$ values were obtained by performing a linear regression over all benchmark scores for that feature-QPU pair (see Fig.~\ref{fig:toronto_correlation} for an example). For each benchmark, the feature is treated as the independent variable and the system performance as the dependent variable.

The two error correction benchmarks (Fig.~\ref{fig:results}c-d) have especially low scores across the majority of the QPUs. This is likely due to the costly \textsc{reset} instructions used in the bit and phase code benchmarks. Indeed, in Fig.~\ref{fig:heatmap} the measurement feature has the strongest correlation with performance for most of the superconducting QPUs (IBM-Lagos-7Q is the exception). For superconducting devices the measurement and reset operations are relatively long compared to the coherence time of the qubits, and so the information stored within the data qubits quickly begins to decay as the number of error correction rounds increases. In contrast, the readout times for trapped ion devices (despite being many times longer than superconducting readout times) are short compared to their long coherence times. This allows the data qubits to sit idly within the ion trap, waiting for the ancilla qubits to be measured and reset, without decohering --- resulting in little correlation between the measurement feature and performance. 

The overwhelming impact of mid-circuit measurements on current system performance is revealed in Fig.~\ref{fig:heatmap_noEC} where again the $R^2$ correlation values are plotted, but in this case the results of the bit and phase code benchmarks have been excluded from the linear regression. When ignoring the results of the error correction (EC) benchmarks, we note improved correlation for many of the feature-QPU pairs. Notably, the correlation of the entanglement-ratio and number of 2-qubit gates features is greatly improved. This suggests that, after \textsc{reset} instructions, entangling operations have the largest impact on system performance. Fig.~\ref{fig:toronto_correlation} provides an example of the linear regression performed over the benchmark scores with and without the EC benchmarks. The difficulty of successfully executing the \textsc{reset} instructions can be seen as the EC benchmarks have significantly lower scores than expected given the value of their entanglement-ratio features.

\section{Discussion}\label{sec:discussion}
The benchmark results presented in Sec.~\ref{sec:results} reveal the variety of tradeoffs that are available to QC system designers, and indicate that competitive advantages can be found by focusing on applications which play to a system's strengths (e.g., faster gate speeds, higher fidelities, denser connectivity). For example, the IonQ device is able to make up for lower two-qubit gate fidelities with better connectivity while the superconducting systems with sparser connectivities are still competitive due to their higher fidelity entangling gates. 

The correlation results in Fig.~\ref{fig:correlations} are a step towards quantitative profiling of quantum programs. In particular, the measurement feature highlights the outsized impact of error correction routines on current system performance. The design of future NISQ systems must focus on improving these operations as mid-circuit measurements are a critical component of quantum error-correcting codes.

Each benchmark was evaluated multiple times to discern the mean system performance. This is partly due to (1) time-variations in the calibrations and fidelities of individual gate operations and (2) the ability of the compiler to find good qubit mappings. The qubit mapping selected by the compiler and the subsequent number of swap insertions has a significant impact on performance since two-qubit gates are so costly. This is evident in the increased variability seen across the superconducting QPUs between the Vanilla QAOA (Fig.~\ref{fig:qaoavanilla-scores}) and ZZ-SWAP QAOA (Fig.~\ref{fig:qaoafswap-scores}) benchmarks. Both benchmarks target the same task, but the all-to-all connectivity of the Vanilla ansatz does not readily match the nearest neighbor connectivity of the superconducting systems. This mismatch is resolved by the compiler which determines a routing schedule among the qubits; a step which introduces extra variability in the performance. Even systems with superior gate fidelities can be severely hampered by sub-optimal compilation. This is especially relevant today when the most popular mode of access is based on a cloud-compute model and the programmer generally does not have total control over the compilation process. A closer investigation of the relationship between compilation and benchmark performance is an important area of future work.

Cloud-based access models also impact our ability to evaluate full variational applications. If the classical and quantum processors are not tightly coupled, then the latency incurred by queue wait times makes the evaluation of variational algorithms with more than 10s of iterations impractical. Systems which support this hybrid quantum-classical programming model are only just starting to appear~\cite{faro2021runtime}. The adoption and availability of this programming model will be crucial for the benchmarking of full variational algorithms.

The cost of collecting the benchmark results presented in this paper influenced our decision to restrict the number of shots per benchmark for the IonQ device. Any quantum benchmark suite will need to be repeatedly evaluated to track the performance of quantum computers over time. The cost of running these benchmarks incentivizes the construction of benchmark suites that provide maximum coverage with as few applications as possible.

\section{Conclusion \& Outlook}\label{sec:conclusion}
SupermarQ is a constantly evolving benchmark suite that adjusts to the fluctuating QC landscape, and it is built with scalability in mind to match the qubit counts of future devices. The included benchmarks are based on real-world applications which makes the suite meaningful to a broad range of use cases, and it provides superior coverage of the application space compared to prior suites and those built entirely from synthetic applications. We plan to open source SupermarQ, which will enable community contributions of additional benchmarks to keep pace with emerging applications.


Computer architects have always been on the forefront of benchmark development for emerging technologies. The SupermarQ suite was inspired by previous work aimed at benchmarking newly emerging computational paradigms like high-performance computers, chip multi-processors, and machine learning systems. Quantum computing's pace of development is currently on an exponential trajectory which has led to varying degrees of skepticism, excitement, and hype. The only way to cut through the hype and accurately ascertain the capabilities of this emerging technology is by returning to the principled, systems-based approach to benchmarking that is at the foundation of computer architecture.

\section*{Acknowledgment}
We would like to thank Akel Hashim for his invaluable help collecting the benchmark results on the AQT device.
This material is based upon work supported by the U.S. Department of Energy, Office of Science, National Quantum Information Science Research Centers, Co-design Center for Quantum Advantage (C2QA) under contract number DE-SC0012704.
This work is also funded in part by EPiQC, an NSF Expedition
in Computing, under grants CCF-1730082/1730449; in part
by STAQ under grant NSF Phy-1818914; in part by NSF
Grant No. 2110860; in part by the US Department of Energy Office 
of Advanced Scientific Computing Research, Accelerated 
Research for Quantum Computing Program; and in part by 
NSF OMA-2016136 and in part based upon work supported by the 
U.S. Department of Energy, Office of Science, National Quantum 
Information Science Research Centers; and supported in part by the National Science Foundation under Grant \# 2030859 to the Computing Research Association for the CIFellows Project; and by DOE grants DE-SC0020289 and DE-SC0020331; and NSF CNS-1763743. 
GSR is supported as a Computing Innovation Fellow at the University of Chicago.
KNS is supported by IBM as a Postdoctoral Scholar at the University of Chicago and the Chicago Quantum Exchange. We also acknowledge support from US Department of Energy Office, Advanced Manufacturing Office (CRADA No. 2020-20099.)

This research used resources of the Oak Ridge Leadership Computing Facility, which is a DOE Office of Science User Facility supported under Contract DE-AC05-00OR22725.
We acknowledge the use of IBM Quantum services for this work. The views expressed are those of the authors, and do not reflect the official policy or position of IBM or the IBM Quantum team.

\section*{Conflicts of Interest}
Fred Chong is Chief Scientist at Super.tech and an advisor to Quantum Circuits, Inc.

\appendix
\section{Artifact Appendix}

\subsection{Abstract}
The artifact contains the source code used to generate, evaluate, and compute the score of the benchmarks presented in this paper. Since the benchmarks in this work utilized proprietary quantum hardware that require valid access tokens, this artifact uses circuit simulation in place of real hardware evaluations. Users which have access to different quantum hardware platforms can take the circuits generated within the artifact and manually execute them. The artifact provides a Jupyter notebook, python files, and benchmark data sets to recreate the plots shown in Figures ~\ref{fig:features}, \ref{fig:results}, \ref{fig:correlations}, and \ref{fig:toronto_correlation}.

\subsection{Artifact check-list (meta-information)}

{\small
\begin{itemize}
  \item {\bf Program:} Cirq.
  \item {\bf Run-time environment:} Jupyter kernel.
  \item {\bf Hardware:} 6-Core Intel Core i7.
  \item {\bf Execution:} Quantum circuit simulation.
  \item {\bf Output:} Benchmark performance scores.
  \item {\bf Experiments:} SupermarQ benchmark applications.
  \item {\bf How much disk space required (approximately)?:} 1 GB to store the artifact directory and python virtual environment.
  \item {\bf How much time is needed to prepare workflow (approximately)?:} 10 minutes.
  \item {\bf How much time is needed to complete experiments (approximately)?:} 30 minutes.
  \item {\bf Publicly available?:} Yes.
  \item {\bf Code licenses (if publicly available)?:} Apache 2.0.
  \item {\bf Workflow framework used?:} Jupyter notebook.
  \item {\bf Archived (provide DOI)?:}\\ \texttt{https://doi.org/10.5281/zenodo.5786391}.
\end{itemize}
}

\subsection{Description}

\subsubsection{How to access}
The artifact is available on Zenodo (\texttt{10.5281/zenodo.5786391}). The source code and artifact notebook are zipped within \texttt{supermarq\_hpca\_ae.tgz}. 

\subsubsection{Hardware dependencies}
The results shown in the paper require access to various quantum computers available over the cloud. Since not all users will have the same access, the artifact relies on quantum circuit simulation available through the Cirq SDK. Any system which can run python programs should be able to evaluate the artifact.

\subsubsection{Software dependencies}
The artifact requires the installation of the SupermarQ python package. The dependencies are listed within \texttt{requirements.txt}.

\subsection{Installation}
The \texttt{README.md} contains detailed instructions to install the SupermarQ python package. After downloading the artifact zipfile, and extracting the contents, the SupermarQ package can be installed via:
\begin{minted}{bash}
  # cd SupermarQ_HPCA_Artifact
  # pip install -r requirements.txt
  # pip install -e .
\end{minted}
The user can then open the jupyter lab with the command:
\begin{minted}{bash}
  # jupyter lab
\end{minted}
The file \texttt{HPCA\_Artifact.ipynb} contains an overview of the benchmarks and figures used in this paper.

\subsection{Evaluation and expected results}
The notebook \texttt{HPCA\_Artifact.ipynb} contains examples showing how the SupermarQ benchmarks are generated and how the scores are computed using the results of the circuit executions (in this case obtained via circuit simulation). The simulations within the notebook utilize a noise model with increasing amounts of noise. This is meant to reflect the real-world execution of these benchmarks on NISQ devices, and as the noise increases we expect that the benchmark score will decrease. The notebook is divided into three parts. The first section, \texttt{Benchmarks}, shows how the benchmark circuits are generated and how the scores are evaluated to create Fig.~\ref{fig:results}. The \texttt{Features} section provides examples of the application feature plots shown in Fig.~\ref{fig:features}. Finally, \texttt{Correlations} walks through the process of creating Fig.~\ref{fig:correlations} and \ref{fig:toronto_correlation}. The Python code used to generate the plots in this last section are contained in \texttt{plotting\_functions.py} and the raw data is stored within the \texttt{data} directory.

\subsection{Methodology}

Submission, reviewing and badging methodology:

\begin{itemize}
  \item \url{https://www.acm.org/publications/policies/artifact-review-badging}
  \item \url{http://cTuning.org/ae/submission-20201122.html}
  \item \url{http://cTuning.org/ae/reviewing-20201122.html}
\end{itemize}


\bibliographystyle{unsrt}
\bibliography{main}

\end{document}